# Thermally activated magnetization reversal in a FeCoB nanomagnet. High-precision measurement method of coercive field, delta, retention time and size of nucleation domain

*Vadym Zayets*

National Institute of Advanced Industrial Science and Technology (AIST), Tsukuba, Japan

Abstract:

*Features of thermally-activated magnetization switching have been studied in a FeCoB nanomagnet using the Néel model. A method of a high-precision measurement of the coercive field, retention time, Δ and the size of the switching nucleation domain has been proposed and experimentally demonstrated using a Hall-probe setup. A high measurement precision, repeatability and reliability are the features of the proposed method. The dependency of the parameters of thermally-activated magnetization switching on the gate voltage and the bias current were studied.*

## *I. Introduction.*

A single-domain ferromagnetic particle has two stable magnetization directions along its easy axes. One data bit can be stored in the particle by means of its two stable magnetization directions. When an external magnetic field is applied opposite to the particle magnetization direction, the magnetization may be reversed to be along the magnetic field and the data bit is recorded. The data bit can be stored until the time when the magnetization is reversed again due to a thermal fluctuation. The understanding of the features of magnetic recording and storage are important for a magnetic memory. This paper studies the properties of the thermally activated switching, which define both the recording mechanism and the storage properties of a single-particle magnetic memory.

Figure 1 shows the schematic diagram of a hysteresis loop of a ferromagnetic nanomagnet. It shows the dependence of the nanomagnet magnetization $M$ on the applied magnetic field $H$. The magnetic field is scanned from a negative to positive value and back to negative. In the case of a sufficiently large magnetic field, the magnetization is always aligned along the magnetic field. However, at a smaller field just after a reversal of the external magnetic field, the magnetization does not follow the reversal and remains in the opposite direction to the magnetic field until the magnetic field reaches the threshold field, at which the magnetization is reversed to be again parallel to the external magnetic field. The threshold magnetic field, at which the magnetization is reversed, is called the coercive field $H_c$ (See Fig.1).

The rectangular shape of the hysteresis loop of Fig.1 indicates that there are no static magnetic domains in the nanomagnet. When there are static domains, their domain walls move under the increasing magnetic field. As a result, the magnetization gradually changes, and there is at least a part of the hysteresis loop, where the magnetization is not a constant and changes gradually[1–3]. When the sizes of a nanomagnet are sufficiently small, there are no static domains and there are two possible magnetization- reversal mechanisms. The magnetization can be reversed coherently when the magnetization at each point on nanomagnet rotates in parallel. This mechanism is called the single- domain magnetization reversal. This mechanism is a feature of the smallest nanomagnets. In case of a slightly larger nanomagnet the reversal mechanism becomes different. The magnetization of only a part of nanomagnet is reversed coherently. Immediately after the reversal the domain wall of this region moves expanding the region of magnetization reversal over the whole nanomagnet (See Fig.3). The region of the initial reversal is called the nucleation domain. For both mechanisms, the value of the magnetization of the nanomagnet does not change at any magnetic field. Only the direction of the magnetization is reversed. As a result, the magnetization loops for these mechanisms have a rectangular shape without any gradual parts (Fig.1). In this paper we study only nanomagnets without static domains that have a rectangular-shape hysteresis loop.

A hysteresis loop for the magnetization switching exists for the following reason. The state, in which the magnetization is opposite to the direction of an external magnetic field, is in an unstable equilibrium. The state, in which the magnetization is parallel to the magnetic field, is more energetically favorable. However, there is an energy barrier between the "up" and "down" magnetization states and the magnetization reversal may occur only when the magnetization overcomes the barrier. The assistance of a thermal fluctuation is required in order to overcome the energy barrier. Because of the critical dependence of the reversal event on the existence of a thermal fluctuation, this type of magnetization reversal is called thermally-activated magnetization switching. The properties of the thermally-activated magnetization switching are important for magnetic data recording and magnetic data storage.

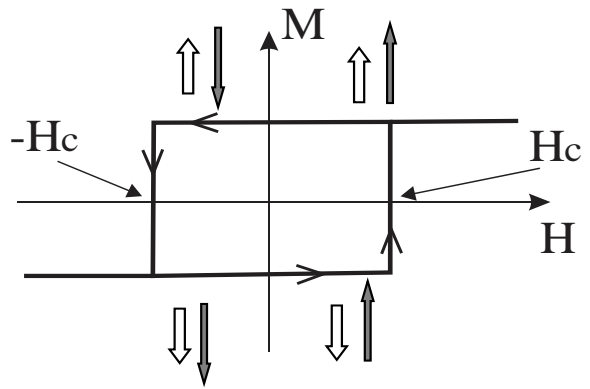

FIG. 1. Schematic diagram of a hysteresis loop of a nanomagnet. The magnetization M of a ferromagnetic nanomagnet as a function of applied external magnetic field $H$. The filled arrow shows the direction of the magnetic field $H$. The unfilled arrow shows the magnetization direction. The nanomagnet has only two "up" and "down" stable magnetization directions along the easy axes. The magnetization switching between the two stable states is abrupt and it occurs at magnetic field H defined as the coercive field $H_c$.

The external magnetic field lowers the height of the energy barrier and makes the probability of a magnetization reversal higher. When the increasing magnetic field reaches $H_c$, the barrier height becomes sufficiently low and the magnetization is reversed. The reduction of the barrier height is not the only method to reverse the magnetization. Even in the case of a higher barrier, the reversal event may occur if the waiting time is longer. A thermal fluctuation of a higher energy is required to overcome a higher energy barrier. Waiting for a longer time makes the probability of the required higher-energy fluctuation greater. Methods of either applying a stronger magnetic field or waiting a longer time both lead to the magnetization reversal. For example, the reversal probabilities may be equal for the cases when the field is smaller but the waiting time is longer or when the field is larger but the waiting time is shorter. An important feature of the coercive field $H_c$ is that it depends on the measurement (waiting) time. For example, the coercive field $H_c$ in Fig.1 becomes larger when the scanning rate of the magnetic field is faster. Since the field, at which the magnetization is switched, depends on the measurement (waiting) time, any definition of the coercive field $H_c$ should include a fixed measurement time (see below).

The dependence of the field, at which the magnetization is switched, on the measurement time has an important implication for any measurement of parameters the thermally –activated switching. The time dependence should be included into any such measurement. Otherwise, a systematic error in the measurement is possible (See Appendix 3). In the proposed measurement method, the time duration, after which the magnetization is switched after applying a magnetic field, is measured as a function of the applied magnetic field. For all our samples, the measured dependence of logarithm of the switching time vs the magnetic field is linear. From this measured linear dependence the parameters of the thermally –activated switching, such as $H_c$, parameter delta $\Delta$, retention time $\tau_{retention}$, and size of a nucleation domain are evaluated. The direct measurement of the time dependence of the switching allows to avoid any systematic error due to the time-dependent features of the thermally –activated switching. An additional merit of the proposed method is a substantially higher measurement precision of $H_c$, $\Delta$ and $\tau_{retention}$ in comparison to other used measurement methods (See Appendix 3). The merits of the measurement of the time dependence of magnetization switching for the evaluation of the $H_c$, and $\tau_{retention}$ have been demonstrated in a multi-particle magnetic system[4,5].

The model of the thermally-activated switching was proposed by Néel [6] in 1955. The Néel model assumes that the switching between two stable magnetization states occurs when the energy of a thermal fluctuation becomes larger than the energy barrier $E_{barrier}$ between two magnetization states. In the Néel model the probability of a thermal-fluctuation is described by the Boltzmann distribution (See Appendix 1).

The complexity of the magnetization reversal is not included in the Néel model. For example, the spin conservation law requires the participation of a particle with a non-zero spin (a magnon, photon etc) in the magnetization reversal process. The complex dynamics of the magnetization reversal should be described by the Landau-Lifshitz (LL) equations, and a model of the thermally-activated magnetization reversal should be based on the LL equation. Aforementioned requirements are not included in the Néel model. The first model of the thermally-activated magnetization reversal, which is based on the LL equation, was developed by Brown [7]. Brown derived the appropriate Fokker-Planck equation for the distribution function of the magnetization orientations from the LL equations using a random magnetic field with white noise properties. This model is called the Néel –Brown model[8].

The validity and applicability of both the Néel and the Néel -Brown models have been intensively studied [8,9]. It was found that the dynamics of the magnetization reversal becomes important for thermally-activated switching only when external conditions change with a frequency comparable with the frequency $f_{FMR}$ of the ferromagnetic resonance (FMR). For example, it is the case when a magnetic pulse of duration comparable with the FMR period $t_{FMR}$ is applied for magnetization reversal or the nanomagnet is illuminated by microwaves with a frequency close to the $f_{FMR}$. This type of reversal is called the resonance magnetization reversal. This paper studies the magnetization reversal at the conditions far away from the resonance type and the complex features of the resonance type of the reversal will not be discussed.

In the case when the magnetization reversal is not of the resonance type, it is not necessary to use the more complex Néel -Brown model with many additional "free" parameters. The simpler Néel model is sufficient and it fully describes all the features of the magnetization reversal [8]. All our experimental measurements are described perfectly by the simpler Néel model. Our experimental data confirms the validity of the use of the Néel model for the description of the non-resonance magnetization switching.

In Chapter 2, the details of the proposed measurement method are described. In this method the time interval, after which the event of the magnetization reversal occurs, is measured. The method requires the detection of the magnetization switching event. In this paper a measurement of the Hall voltage is used to detect the magnetization direction and consequently a magnetization switching event. The experimental details of implementation of the proposed measurement method with the Hall setup are described in Chapter 3. The application of the proposed method for a measurement of parameters of the thermally- activated switching is described in Chapter 4. The features of the different magnetization switching mechanisms are discussed in Chapter 4. Chapters 5 and 6 describe high-precision measurements of the voltage-control magnetic anisotropy (VCMA) effect and the spin-orbit torque (SOT) effect, correspondingly. In Appendix 1, the Arrhenius law, which describes the average time of the magnetization reversal, is obtained from the Boltzmann energy distribution. In Appendix 2, the measurement method of the linearly-ramp magnetic field is described. In Appendix 3, the proposed method is compared with other used methods for a measurement of the parameters of the thermally- activated magnetization switching. Possible systematic errors and limitations of different measurement methods are discussed.

## II. Measurement of the coercive field and retention time according to the Néel model

In the following, the magnetization switching time $t_{switch}$ is calculated according to the Néel model for a nanomagnet, which has the perpendicular magnetic anisotropy (PMA). The easy axis of the nanomagnet is perpendicular-to-plane. There is a difference of the magnetic energy of the nanomagnet between the in-plane and perpendicular-to-plane magnetization directions. This difference is called the PMA energy. In the case when an external magnetic field H is applied along the easy axes and therefore perpendicularly to the film, the magnetic energy of a nanomagnet can be calculated as [10]:

$$E = -E_{PMA} \cos^2(\theta) + M \cdot H \cdot \cos(\theta) \quad (1)$$

where θ is the angle between the magnetization M and the film normal. $E_{PMA}$ is the PMA energy, which includes the contribution due to the demagnetization field. The first term in Eq.(1) describes the PMA energy and the second term describes the magnetic dipole energy.

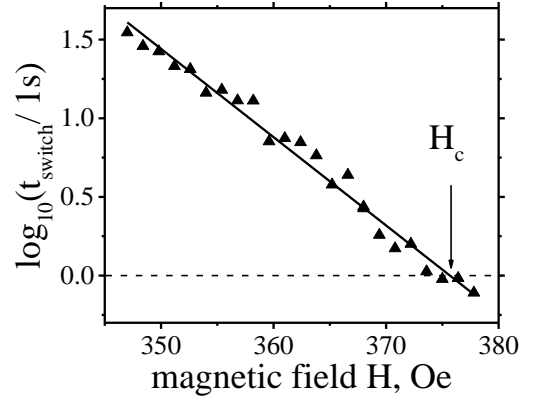

FIG. 2. Measured magnetization reversal time $t_{switch}$ as a function of applied perpendicular magnetic field $H$. The switching time of 1 second corresponds to the coercive field $H_c$.

Eq.(1) has two minimums $E_{min,\uparrow\uparrow}$ and $E_{min,\uparrow\downarrow}$, which correspond to the magnetization direction along and opposite to the magnetic field. The maximum $E_{max}$ of Eq.(1) corresponds to the energy barrier between the ↑↑ and ↑↓ states. From Eq.(1), the energy barrier $E_{barrier}$ is calculated as

$$E_{barrier} = E_{max} - E_{min,\uparrow\downarrow} = E_{PMA}\left(1 - \frac{M \cdot H}{2 \cdot E_{PMA}}\right)^2 \quad (2)$$

The magnetization angle at the barrier position is calculated as

$$\cos(\theta_{barrier}) = \frac{M \cdot H}{2 \cdot E_{PMA}} \quad (3)$$

Eqs. (2),(3) can be simplified using the anisotropy field $H_{anis}$. The $H_{anis}$ is defined as the in-plane magnetic field, at which initially-perpendicular magnetization turns completely into the in-plane direction. Since the measurement of $H_{anis}$ is straightforward and unambiguous (see below), the use of the $H_{anis}$ instead of the $E_{PMA}$ is convenient. $H_{anis}$ can be measured by applying an in-plane magnetic field and monitoring the in-plane component of the magnetization. The relation between $E_{PMA}$ and $H_{anis}$ can be found as follows. In the case when an external magnetic field H is applied in-plane, the magnetic energy of a nanomagnet can be calculated as:

$$E = -E_{PMA} \cos^2(\theta) + M \cdot H \cdot \sin(\theta) \quad (4)$$

The minimum of the energy (Eq.(4)) gives the magnetization angle $\theta_{min}$ as

$$\sin(\theta_{min}) = \frac{M \cdot H}{2 \cdot E_{PMA}} \quad (5)$$

By definition, the magnetization turns fully in-plane at $H=H_{anis}$ when $\theta_{min}=90^0$. At this condition, Eq.(5) gives

$$E_{PMA} = \tfrac{1}{2} H_{anis} M \quad (6)$$

The substitution of Eq. (6) into Eqs. (2) and (3) gives

$$E_{barrier} = E_{PMA}\left(1 - \frac{H}{H_{anis}}\right)^2 \quad (7)$$

$$\cos(\theta_{barrier}) = \frac{H}{H_{anis}} \quad (8)$$

Eq. (7) can be further simplified using the fact that the magnetic field, at which the magnetization is switched, is substantially smaller than the $H_{anis}$:

$$H \ll H_{anis} \quad (9)$$

For example, it was measured that $\frac{H}{H_{anis}} \approx 1\% - 3\%$ for all our samples. As was aforementioned, the switching field becomes larger as the switching time becomes shorter. In our experimental setup, the shortest measurement time is 100 ms and the condition (9) is well satisfied. However, in the case of a shorter measurement time, the condition (9) may be not satisfied. The extension of the reported results for that case is straightforward. Using condition (9) and substituting Eq. (6) into Eq. (7) gives

$$E_{barrier} \approx E_{PMA}\left(1-\frac{2H}{H_{anis}}\right) = E_{PMA} - H \cdot M \qquad (10)$$

An assumption of the Néel model is that the average magnetization reversal time $t_{switch}$ between two stable magnetic states is described by the Arrhenius law (See Appendix 1):

$$t_{switch} = \frac{1}{f_0}\exp\left(\frac{E_{barrier}}{kT}\right) \qquad (11)$$

where $f_0$ is the attempt frequency. Substituting Eq.(10) into Eq.(11) gives the $t_{switch}$ as

$$t_{switch} = \tau_{retention}\exp\left(-\frac{H \cdot M}{kT}\right) \qquad (12)$$

where $\tau_{retention}$ is the retention time and it is defined as the average time, after which the magnetization is reversed in the absence of a magnetic field. Also, $\tau_{retention}$ is referred to the maximum data storage time of a magnetic memory. From Eqs.(11),(12) the $\tau_{retention}$ is calculated as

$$\tau_{retention} = \frac{1}{f_0}\exp\left(\frac{E_{PMA}}{kT}\right) \qquad (13)$$

The Eq.(12) can be rewritten as a linear dependence:

$$\log(t_{switch}) = \log(\tau_{retention}) - \frac{M}{kT} \cdot H = \log(\tau_{retention}) - slope \cdot H \qquad (14)$$

where $slope = M/kT$.

As aforementioned, the coercive field $H_c$ is defined as the perpendicular magnetic field, at which the magnetization reversal occurs. As follows from Eq. (12), $H_c$ has no physical meaning without a specification of the measurement time. The longer the measurement time $t_{switch}$ is, the smaller value is of magnetic field $H$, at which the magnetization is switched. For example, when the measurement time equals to $\tau_{retention}$, the $H_c$ equals to zero. In order to match a commonly-used setup for the measurement of $H_c$, in this paper we refer to all values of $H_c$ for the measurement time of one second:

$$H_c = \frac{\log(\tau_{retention})}{slope} \qquad (15)$$

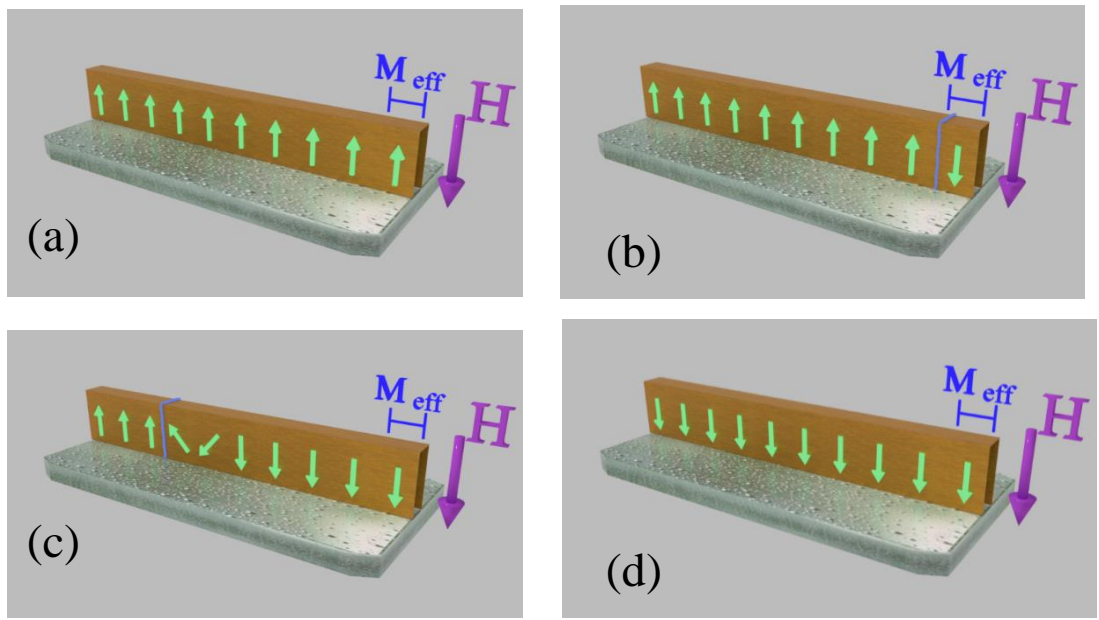

FIG. 3. Magnetization reversal of the nucleation-domain type in a nanomagnet of an elongated shape. Temporal evolution of the reversal dynamic. (a) An external magnetic field $H$ is applied opposite to the magnetization of the nanomagnet (green arrows); (b) magnetization $M_{eff}$ of the nucleation domain is reversed to be along $H$; (c) the domain wall (blue line) of nucleation domain moves along the nanowire expanding its size; (d) the magnetization of the whole nanowire is reversed to be along H.

The magnetization switching time was measured as follows. At first, the magnetization direction is aligned along the easy axis by applying a large reset magnetic field $H_{reset}$. Next, the magnetic field H was applied opposite to the magnetization direction and $H_{reset}$. The time interval $t_{switch}$, after which the magnetization is reversed, was measured. The measurement procedure was repeated 200 times for each value of H and a statistical analysis was applied to find the average of $t_{switch}$. The $t_{switch}$ exponentially increases when the magnetic field H increases. Figure 2 shows the measured $t_{switch}$ as a function of the H. On a logarithmic scale, the $t_{switch}$ is linearly proportional to the magnetic field as it predicted by the Néel model (Eq.(14)).

The extrapolation of the line till the intersection with the y-axis gives the retention time $\tau_{retention}$. For a nanomagnet of a different size and thickness, the $\tau_{retention}$ was measured to vary between a few minutes (a soft-magnetic sample) and a billion years (a hard-magnetic sample). The intersection of the line and the x-axis gives the coercive field $H_c$. From the statistical analysis, the measurement precision of $H_c$ is high and estimated to be about 0.6 Oe.

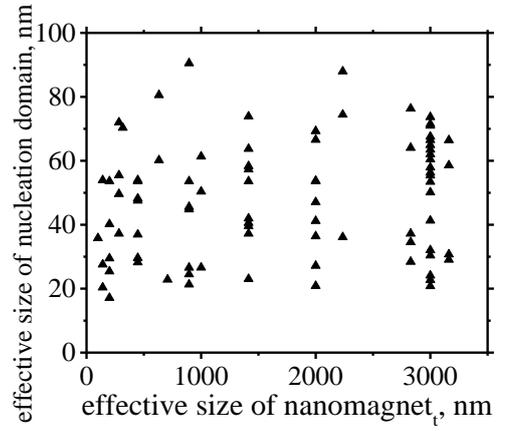

FIG. 4. The effective size of the nucleation domain for magnetization reversal vs the effective size of the nanomagnet. Each point corresponds to a nanomagnet of a different size. The effective sizes of domain and nanomagnet equal to the square root of their area. The area of the nucleation domain is evaluated from the data of Fig.2 and Eq. (15).

### III. Experimental details

In order to measure the switching time $t_{switch}$ of a nanomagnet, the monitoring of its magnetization direction is required. It can be done using the dependence of either magneto-resistance or the Hall angle or the Kerr rotation angle on the magnetization direction. In the paper, all measurements are done using the Anomalous Hall effect (AHE). The AHE configuration has several advantages compared to the configuration based on the magnetic tunnel junction (MTJ). Firstly, there is no undesirable influence of the dipole magnetic field from the reference electrode on measured properties and there is no undesirable influence of the spin transfer torque due to the flow of the spin-polarized current from the reference electrode. Secondly, different materials of the gate electrode can be tested. The MTJ configuration is limited to a specific ferromagnetic metal, which has to provide a sufficient TMR. There is no such limitation for the AHE configuration.

The samples were fabricated on a Si/SiO2 substrate by sputtering. The layer stack is $SiO_2$:NM FM:MgO:$SiO_2$. Ta and W were used as the nonmagnetic metal (NM). FeB and FeCoB were used as the ferromagnetic metal (FM). A variety of samples with different thicknesses of NM in the range of 2 to 5 nm and the thicknesses of FM in the range of 0.7 to 1.4 nm were fabricated and measured. The samples with a thicker FM are magnetically softer and have a smaller $H_c$ and $H_{anis}$. Nanowires of a different width between 100 and 1000 nm and of a different length between 50 nm to 1000 nm were fabricated by the argon milling. The FeB or FeCoB layers were etched out from the top of the nanowire except for a small region of the nanomagnet, which was aligned to the Hall probe. The width of the Hall probe was 50 nm. The shape of measured hysteresis loop was rectangular indicating the absence of static domains.

### IV. Comparison of magnetization reversal mechanisms. Measurement of $\Delta$ and size of nucleation domain

There are static magnetic domains in a ferromagnetic material of a large volume. The size and shape of the static magnetic domains is determined by a balance between the magnetostatic energy and exchange energy (the energy of a domain wall). The magnetization reversal mechanism in a material, that has static domains, is due to the domain wall motion and the domain expansion. The size and distribution of the static magnetic domain depends on the applied external magnetic field. As the magnetic field increases, the volume of the domains of magnetization parallel to the field gradually becomes larger and the volume of the domains of magnetization opposite to the field becomes smaller. As a result, the total magnetization of the ferromagnetic material increases gradually as the magnetic field increases. The part of hysteresis loop, where the magnetization is not a constant and changes gradually, is an indication of existence and expansion of the static domains.

The magnetostatic energy decreases as the volume of ferromagnetic material decreases. When the volume becomes sufficiently small, the magnetostatic energy is unable to balance the energy of a domain wall, a static magnetic domain cannot be created and the magnetization of the whole volume of a ferromagnetic material is aligned in one direction. Such ferromagnetic material is defined as a nanomagnet. There are two possible mechanisms for magnetization reversal of the nanomagnet: the single-domain reversal and the nucleation- domain reversal. For both reversal mechanisms, the shape of the hysteresis loop is rectangular (Fig.1).

When the dimensions of a nanomagnet are sufficiently small, the magnetization reversal occurs in a single domain. It means that the magnetization at all points of the nanomagnet rotates coherently and the magnetization in different parts of the nanomagnet remains parallel during the rotation. When the dimensions of the nanomagnet become slightly larger, the type of the magnetization reversal is changed to the nucleation-domain type (see Fig.3). In the case of the larger nanomagnet, it is more energy favorable when at first the magnetization of only a small nucleation domain is reversed following by domain wall movement expending the region of the reversed magnetization over the whole nanomagnet. The difference between the static and nucleation domains is that the nucleation domain is unstable. In a nanomagnet, the magnetostatic energy is small and unable to balance the energy of a domain wall. Therefore, as soon as a nucleation domain is created, it immediately expands over the whole nanomagnet. The existence and expansion of the nucleation domain during the magnetization reversal of a nanomagnet were directly observed by a time-resolved measurement of the X-ray magnetic circular dichroism (XMCD)[11].

The following explains the reason why the magnetization reversal of the nucleation-domain type exists. From consideration of the total energy, the single-domain mechanism is always preferable. The total magnetic energy of the single-domain state is smaller than the total energy of the state with a nucleation domain. It is the reason why the nucleation domain is unstable and immediately expands into the single-domain state. However, the energy barrier for the magnetization reversal by the nucleation- domain mechanism is smaller and it governs the reversal process. The energy barrier between two stable magnetization states is proportional to the magnetostatic energy (See Eq. (7)), which does not include the energy of the domain wall. As a result, the energy barrier is smaller for a smaller volume of a magnetic material. Magnetization reversal occurs only when the energy of a thermal fluctuation is comparable with the energy barrier between two stable states of the opposite magnetization. Since the volume of nucleation domain is smaller than the volume of the whole nanomagnet, the required energy of a thermal fluctuation is smaller to reverse the magnetization of the nucleation domain than to reverse the magnetization of the whole nanomagnet.

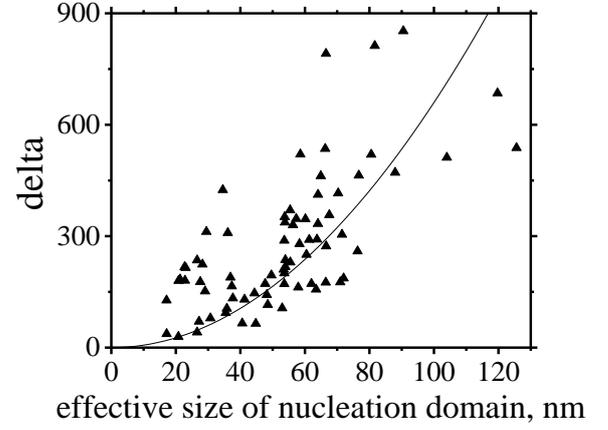

FIG. 5. Measured $\Delta$ vs the effective size of the nucleation domain. Each point corresponds to a different nanomagnet of a different size. Solid line shows a fit assuming the linear dependence of $\Delta$ on the volume of the nanomagnet. The data of magnetically-harder samples are above the line. The data of softer samples are below the line (See Eqs. (15),(17)).

The size of the nucleation domain cannot be infinitely small. The total energy of the nanomagnet with a nucleation domain increases when the size of the nucleation domain decreases, because of the reduction of the magnetostatic energy. There is an optimum size of the nucleation domain when the energy barrier is lower, but the increase of the magnetic energy is still moderate. Additionally, the length, position and energy of the domain wall should influence the optimum size of the nucleation domain. When the size of the nanomagnet becomes smaller than the size of the nucleation domain, the magnetization switching mechanism changes to the single-domain type.

As can be seen from Eq.(3), the energy barrier can be directly evaluated from the measurement of Fig.2. The exponential decrease of the switching time is due to the existence of the energy barrier, which can be evaluated by fitting. From Eq.(1)), the energy barrier is linearly proportional to the magnetization $M$, which is reversed by a thermal fluctuation. The magnetization $M$ corresponds to the magnetization of nucleation domain $M_{domain}$ in the case of nucleation-domain reversal and the magnetization of the nanomagnet $M_{magnet}$ in the case single-domain reversal. From Eq. (12), the magnetization $M$ is linearly proportional the slope of the line of Fig.2. Knowing the magnetization $M_{domain}$ of the nucleation domain, it is possible to calculate its volume. Since the total magnetization of a ferromagnetic region is proportional to its volume, the volume of the nucleation domain $V_{domain}$ is calculated as:

$$V_{domain} = V_{magnet} \frac{M_{domain}}{M_{magnet}} \qquad (16)$$

where $V_{magnet}$ is the volume of the nanomagnet. Substitution Eq.(12) into Eq.(14) at T=300 K gives

$$V_{domain}\left(nm^3\right) = 51717 \cdot \frac{slope\left(Gauss^{-1}\right)}{M_{ferro}/V\left(Tesla\right)} \qquad (17)$$

where *slope* is the slope of the line of Fig.2. $M_{ferro}/V$ is the magnetization per volume of the ferromagnetic material of the nanomagnet in Teslas, which can be measured by a magnetometer. For the single-domain reversal, $V_{domain}=V_{magnet}$.

Figure 4 shows the evaluated effective size of the nucleation domain as a function of the effective size of the nanomagnet. The effective sizes, which are defined as the square root of area of a nucleation domain or a nanomagnet, are used for the following reason. Nanomagnets of different shapes were used for the measurement. The exact shape of nucleation domain is unknown. To compare the data from nanomagnets and domains of a variety of different shapes, the effective size is descriptive.

The data of Fig.4 indicates that the size of the nucleation domain does not clearly depend on the size of the nanomagnet for a larger nanomagnet (effective size >200 nm) and is slightly reduced for a smaller nanomagnet (effective size <200 nm) There is a substantial variation of the domain sizes in the range between 20 nm and 90 nm. This means that the domain size substantially depends on the nucleation site. The existence of a fabrication defect, or a material defect, or a geometrical defect, or a structure imperfection must substantially influence the size of the nucleation domain. When the size of the nucleation domain decreases, the slope of the line of Fig.2 decreases (Eq.(15)), the retention time $\tau_{retention}$ decreases and the coercive field $H_c$ increases. Therefore, the size of the nucleation domain is a critical parameter, which determines the major magnetic properties of a nanomagnet.

For a magnetic memory application, it is important to identify the critical nanomagnet size, at which the magnetization reversal mechanism changes from the nucleation- domain type to the single-domain type. The variation of switching parameters between memory cells of one memory chip is smaller when the switching type of all memory cells is the single-domain type. From Fig.4, this critical size depends not only on the nanomagnet material, but also on the fabrication technology.

The parameter $\Delta$ is a parameter, which estimates the ability of a memory cell to withstand a thermal fluctuation and the ability to withstand the temperature rise without loss of the stored data. It is defined as the ratio of energy barrier $E_{barrier}$ in the absence of an external magnetic field to the thermal energy $kT$. From Eqs.(6), (13)-(14), $\Delta$ can be calculated as

$$\tau_{retention} = \frac{1}{f_0} e^{\Delta}$$

$$\Delta = \frac{E_{barrier}(H=0)}{kT} = \frac{E_{PMA,}}{kT} = \frac{M \cdot H_{anis}}{2kT}$$

(18)

It should be noted that $\Delta$ decreases with a rise of temperature (See Eq.(13)). The substitution Eq.(14) into Eq.(18) gives

$$\Delta = 0.5 \cdot slope(Gauss^{-1}) \cdot H_{anis}(Gauss) \quad (19)$$

where *slope* is the slope of line of Fig.2. As was aforementioned, $H_{anis}$ can be measured by applying an in-plane magnetic field and monitoring the in-plane component of magnetization. The $H_{anis}$ is the field, at which the magnetization turns fully in-plane. Substitution of Eq.(15) into Eq.(19) gives

$$\Delta = 0.5 \cdot \log(\tau_{retention}) \cdot \frac{H_{anis}}{H_c} \quad (20)$$

From Eq.(20), both the parameter $\Delta$ and log ($\tau_{reten}$) may be equally used to characterized the performance of a magnetic memory, because of their linear relation.

Figure 5 shows the dependence of $\Delta$, which was evaluated from Eq.(19), on the size of nucleation domain, which was evaluated from Eq.(17). Each point corresponds to a measurement of a nanomagnet of a different size and thickness. From Eq.(18), $\Delta$ should be linearly proportional to the magnetization of the nucleation domain $M$ and therefore the volume of the nucleation domain. The data of Fig.5 confirms such dependence. However, some data deviates from the line. This is due to the variation of $E_{PMA}$ (or $H_{anis}$) from a nanomagnet to a nanomagnet (compare Eqs. (17) and (19)).

### V. The VCMA effect

The VCMA effect describes the fact that in a capacitor, in which one of the electrodes is made of a thin ferromagnetic metal, the magnetic properties of the ferromagnetic metal are changed, when a voltage is applied to the capacitor.

The dependency of the coercive field on the gate voltage was reported for different materials. However, there are contradictions in the reported polarity of such a dependency. Both the negative and positive slopes were reported for dependency of $H_c$ on the gate voltage. A linear voltage dependency of the coercive field with a negative slope was measured in Ta:Fe$_{0.4}$Co$_{0.4}$B$_{0.2}$ [12,13], Au:Fe$_{80}$Co$_{20}$ [14], Ru:Co$_2$FeAl[15] and with a positive slope in Pd:FePd [16], Ta:Fe$_{0.4}$Co$_{0.4}$B$_{0.2}$ [17], Pt:Co[18,19]. As it is shown below (Fig.6(b)), the change of $H_c$ is relatively small under a gate voltage. An insufficient precision of a $H_c$ measurement might be the reason for the contradictions of the reported polarity of the dependency of $H_c$ on the gate voltage.

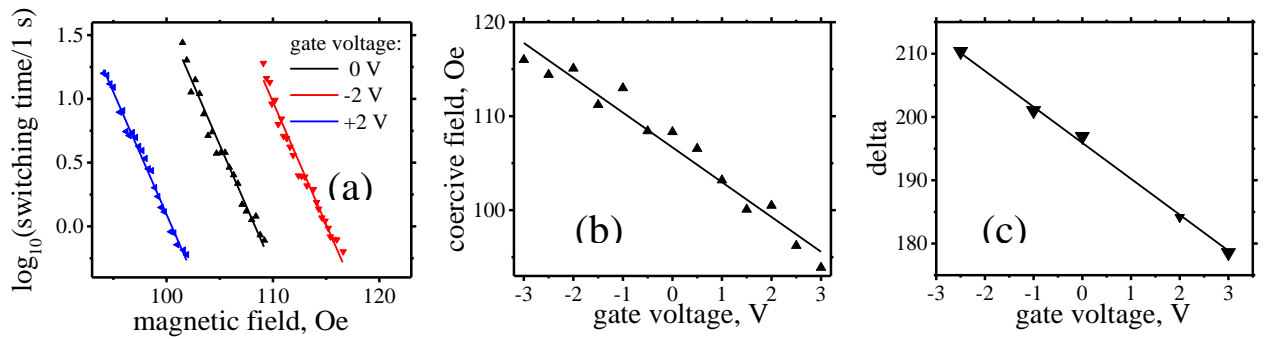

FIG. 6. (a) Magnetization switching time $t_{switch}$ as a function of the external magnetic field at gate voltage of -2, 0, +2 V. (b). The coercive field $H_c$ vs the gate voltage. (c) $\Delta$ as a function of the gate voltage.

For our VCMA measurements, a Ta/Ru metal gate electrode was fabricated on the top of the MgO instead of the SiO2 cover[20–22]. The positive gate voltage means that a positive voltage was applied to the non-magnetic gate electrode. In order to increase the break-down voltage and to suppress the oxygen diffusion in the gate the following growth procedure was used. At first, a 1 nm MgO was deposited at room temperature. Next, the sample was annealed at $220^0$ C for 30 minutes and the remaining of the MgO gate oxide was grown at $220^0$ C. Three 30-minute growth interruptions after each 1 nm of growth were used to improve the MgO crystal quality.

Figure 6 (a) shows the dependence of $t_{switch}$ on the applied perpendicular magnetic field $H$, which is measured at a different gate voltage. The $t_{switch}$, $\tau_{retention}$ and $H_c$ increase at a negative gate voltage and decrease at a positive gate voltage. The slope of the lines does not depend on the gate voltage. It means that the size of the nucleation domain is independent on the gate voltage (Eq.(17)). Figures 6(b) and 6(c) show the measured $H_c$ and $\Delta$ as a function of the gate voltage. All measurement points fit well into a straight line and the polarity is negative. It should be noted that the measured gate-voltage dependencies of $H_{anis}$, $E_{PMA}$, the Hall voltage[20] and the spin polarization[23], all have a similar negative slope, which might be a feature of the VCMA effect[20]. The measured slope of the gate-voltage dependency of $H_c$ was the same for "up to down" magnetization switching (left slope of the loop of Fig.1) and for "down to up" magnetization switching (right slope of the loop of Fig.1).

## VI. The SOT effect

The SOT effect describes the fact that magnetic properties of ferromagnetic nanowire may depend on the magnitude and polarity of an electrical current flowing through the nanowire[24,25]. For example, under a sufficiently large current the magnetization of the nanowire may be reversed[26–28]. The direction of the magnetization reversal depends on the polarity of the current. The effect may be used as a recording mechanism for the 3-terminal MRAM[29,30].

The parameters of the thermally-activated switching are affected by the polarity and the magnitude of the bias current due to the SOT effect as well. The dependence of $H_c$ on the current density and current polarity was reported in Pt:Co:AlO$_x$ [28], Ta:Co:TaO$_x$[27], IrMn:CoB:Pt[31]

Figure 7(a) shows the dependence of $t_{switch}$ on $H$ measured at a different density and polarity of the bias current. $t_{switch}$, $\tau_{retention}$ and $H_c$ decrease when the bias current increases. Additionally, $t_{switch}$, $\tau_{retention}$ and $H_c$ depend on the polarity of the bias current. The slope of the lines depends on the current density and current polarity. The slopes are larger at a negative current and smaller at a positive current. This means that the size of the nucleation domain depends on the current density and current polarity. This effect can be understood as follows. An electrical current induces the spin transfer torque, under which a domain wall may move. As a result, the size of nucleation domain for the magnetization reversal becomes current-dependent. The domain wall movement under a bias current is the well- studied effect for a static domain. For a nucleation domain, the observation of the domain- wall motion is a challenging task, because the nucleation domain is unstable and exists for a short time. The dependency of the size of the nucleation domain on the bias current due to the SOT effect was observed by a time-resolved measurement of the XMCD[11].

Figure 7(b) shows the measured $H_c$ vs. the current density. $H_c$ decreases for both polarities of the current. The used current density is relatively large and the decrease of $H_c$ is assumed to be due to the heating of the nanomagnet. The increase of the nanowire resistance confirms the increase of the nanowire temperature. In order to exclude the influence of heating, $H_c$ was measured at the same current, but for two opposite current directions. Figure 7(c) shows the change of the spin polarization as the polarity is reversed. Each point corresponds to the difference of $H_c$ for two opposite current polarities measured at the same current. Data of positive and negative currents correspond to two separate measurements made at the same conditions. The change of $H_c$ depends linearly on the current. The slope of the line is opposite for the magnetization reversal from spin-up to spin-down direction (switching at a negative H of Fig.1) and for reversal from the spin-down to spin-up direction (switching at a positive H of Fig.1).

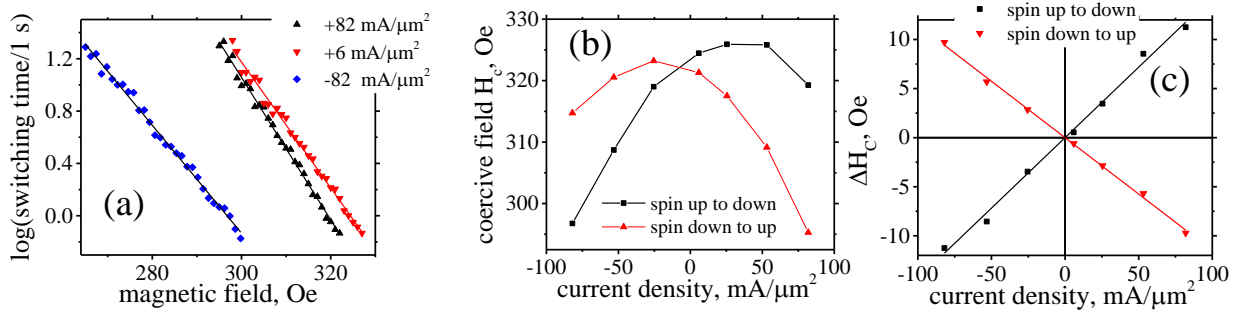

FIG. 7. (a) Magnetization switching time $t_{switch}$ as a function of external magnetic field at a different density and polarity of the bias current. (b). Coercive field $H_c$ vs. density of the bias current. (c) Difference of $H_c$ under reversal of current polarity vs. density of the bias current.

For an effective operation of MRAM based on the SOT effect, the retention time $\tau_{retention}$ should be shortest during a recording event and longest during a data storage. For the nanomagnet of Fig. 7(a), the negative bias current reduces $\tau_{retention}$ significantly from $10^{17.38}$ seconds=7.6E9 years to $10^{12.24}$ seconds=5.5E4 years.

## *VII. Conclusion*

A method of a high-precision measurement of the retention time $t_{retention}$, coercive field $H_c$, parameter $\Delta$ and the size of a nucleation domain for magnetization reversal has been proposed and demonstrated. The measurement precision of 0.6 Oe for $H_c$ has been demonstrated. The measurement precision of about 2% for $\Delta$ and 3 nm for size of a nucleation domain has been achieved.

It has been demonstrated that in the case when the magnetization switching time is substantially longer than $t_{FMR}$ and the frequency of alternation of the magnetic field is substantially slower than the FMR frequency, the classical Néel model [6] perfectly describes all the features of the thermally-activated magnetization switching. The main prediction of the classical Néel model is that the logarithm of the magnetization switching time is linear proportional to the energy barrier between stable magnetization state (Eq.11) and therefore linearly proportional to the external magnetic field (Eq.12). All reported experimental measurements perfectly fit to this linear dependence (Fig.2, Fig.6a, Fig.7a). It clearly proves that the classical Néel model fully describes the non-resonance magnetization reversal. It has several consequences. One of them is that there are only two independent parameters for the magnetization switching of a nanomagnet. This fact allows avoiding many possible systematic errors, which are very common for a measurement of thermally-activated magnetization switching (Appendix 3).

The dependencies of the parameters of thermally-activated magnetization switching on the gate voltage (the VCMA effect) and the bias current (the SOT effect) have been systematically studied. The change of the parameters under a gate voltage or due to the reversal of the polarity of the bias current is small. The high measurement precision of the proposed method is beneficial for the study of these small changes.

In the case of the VCMA effect, $H_c$, retention time, magnetization switching time and $\Delta$ linearly increase under a negative gate voltage and decrease under a positive gate voltage. From the same-polarity linear dependence on the gate voltage has been measured for the anisotropic field $H_{anis}$, the Hall angle[20] and the spin polarization[23], it may be suggested that such gate-voltage dependency is a feature of the VCMA effect.

In the case of the SOT effect, $H_c$, $\tau_{retention}$, $t_{switch}$, $\Delta$ and the size of the nucleation domain depend on the polarity and the magnitude of the bias current. There is a difference of $H_c$ when the polarity of the bias current is reversed. The difference depends linearly on the magnitude of the bias current. The polarity of the dependence is opposite for cases when the magnetization switches from the spin-up to the spin-down state or it switches in the opposite direction.

A method to measure of the size of nucleation domain for the magnetization reversal has been developed and demonstrated. The magnetization reversal mechanism of a nanomagnet can either be of the single-domain type or the nucleation- domain type. In the case of a larger nanomagnet, the mechanism of the magnetization reversal is of the nucleation- domain type. For this mechanism, at first the magnetization of only a small nucleation domain is reversed following by the domain wall movement expending the region of the reversed magnetization over whole nanomagnet. The size of the nucleation domain is evaluated from the magnetization-switching data of Fig.2. In the case of a CoFeB nanomagnet, the average size of the nucleation domain is 50 nm. The size of the nucleation domain does not depend on the gate voltage, but it does depend on the polarity and magnitude of the bias current.

## *Appendix 1*

In the following, the Arrhenius law (Eq.(11)), which calculates the magnetization switching time $t_{switch}$, is obtained from the Boltzmann energy distribution. The Néel model assumes that the magnetization reversal occurs only when the spin of the nanomagnet interacts with a "non-zero-spin" particle (a magnon, a photon etc.), which energy is higher than the barrier height $E_{barrier}$ between two stable states of the nanomagnet. The temperature is assumed to be sufficiently high so that the energy distribution of the particles is described by the Boltzmann distribution. Therefore, the number of particles, which are able to reverse the magnetization, is calculated as

$$n_{rever} = n_0 \exp\left(-\frac{E_{barrier}}{kT}\right) \tag{A1.1}$$

where $n_0$ is the total number of the particles, which are able to reverse the magnetization.
The frequency, at which the magnetization can be reversed, is calculated as

$$f_{rever} = f_{inter} \cdot n_{rever} = f_{inter} \cdot n_0 \exp\left(-\frac{E_{barrier}}{kT}\right) \tag{A1.2}$$

where $f_{inter}$ is the frequency of interaction of one particle with the spin of the nanomagnet.
The probability $P_{rever}(t,t+dt)$ of the magnetization reversal in a small time interval between t and t+dt is calculated as

$$P_{rever}(t,t+dt) = f_{rever} \cdot dt = f_{inter} \cdot n_0 \exp\left(-\frac{E_{barrier}}{kT}\right) \cdot dt = \frac{dt}{\tau} \tag{A1.3}$$

where $\tau = \frac{1}{f_{inter} \cdot n_0} \exp\left(\frac{E_{barrier}}{kT}\right) = \tau_0 \exp\left(\frac{E_{barrier}}{kT}\right) \tag{A1.4}$

The probability $P_{not}(t,t+dt)$, that the magnetization is not reversed during the time interval dt, is calculated is

$$P_{not}(t,t+dt) = 1 - P_{rever}(t,t+dt) = 1 - \frac{dt}{\tau} \tag{A1.5}$$

If the magnetization is not reversed in the interval $[t_0,t+dt]$, that means that it is not reversed in both intervals $[t_0,t]$ and $[t,t+dt]$. Therefore, the probability $P_{not}(t_0,t+dt)$ is calculated as

$$P_{not}(t_0,t+dt) = P_{not}(t_0,t) \cdot P_{not}(t,t+dt) = P_{not}(t_0,t) \cdot \left(1 - \frac{dt}{\tau}\right) \tag{A1.6}$$

Eq.(A1.6) can be simplified as

$$dP_{not} = P_{not}(t_0,t+dt) - P_{not}(t_0,t) = -P_{not}(t_0,t)\frac{dt}{\tau} \tag{A1.7}$$

The function $P_{not}(t)$ is defined as the probability of non-reversal of the magnetization in the time interval from $t_0$ to $t$. From Eq. (A1.7), the function $P_{not}(t)$ satisfies the following differential equation:

$$\frac{dP_{not}}{P_{not}} = -\frac{dt}{\tau} \tag{A1.8}$$

In the case when the external magnetic field H is a constant and time-independent, the energy barrier $E_{barrier}$ and $\tau$ are time-independent as well and Eq. (A1.8) becomes a linear differential equation. The solution of Eq.(A1.8) gives the probability $P_{not}(t)$ of the non-reversal of the magnetization in the time interval from $t_0$ to $t$ as:

$$P_{not}(t) = \exp\left(-\frac{t-t_0}{\tau}\right) \tag{A1.9}$$

Next, the averaging magnetization switching time $t_{switch}$ is calculated. If the external magnetic field is switched on at time $t_0=0$ in the direction opposite to the magnetization, the probability $dp_{switch}$ that the magnetization is reversed in the time interval between $t$ and $t+dt$ is the difference between probabilities that it is not reversed until time $t$ and until time $t+dt$:

$$dp_{switch}(t) = P_{not}(t) - P_{not}(t+dt) = \exp\left(-\frac{t}{\tau}\right) - \exp\left(-\frac{t+dt}{\tau}\right) =$$
$$= \exp\left(-\frac{t}{\tau}\right)\left[1 - \exp\left(-\frac{dt}{\tau}\right)\right] = \exp\left(-\frac{t}{\tau}\right) \cdot \frac{dt}{\tau} \tag{A1.10}$$

From Eq. (A(11.10), the averaging magnetization switching time $t_{switch}$ is calculated as

$$t_{switch} = \int_{t=0}^{\infty} t \cdot dp_{switch} = \tau \tag{A1.11}$$

The substitution of Eq.(A1.4) into (A1.11) gives the Arrhenius law (Eq.(11)) as

$$t_{switch} = \tau_0 \exp\left(\frac{E_{barrier}}{kT}\right) = \frac{1}{f} \exp\left(\frac{E_{barrier}}{kT}\right) \tag{A1.12}$$

## Appendix 2

In the following, the probability of the thermal-activated switching is calculated in the case when the applied external magnetic field changes over time. In this case, the height of the energy barrier $E_{barrier}$ becomes time-dependent and the solution of the differential equation Eq. (A1.8) is

$$\ln\left[P_{not}(t)\right] = -\int_0^t \frac{d\tilde{t}}{\tau(\tilde{t})} = \frac{-1}{\tau_0} \int_0^t \exp\left(-\frac{E_{barrier}(\tilde{t})}{kT}\right) \cdot d\tilde{t} \tag{A2.1}$$

The linearly-ramped magnetic field can be described as

$$H(t) = H_0 \frac{t}{t_0} \tag{A2.2}$$

The substitution of Eq. (A2.2) into Eqs.(7) and (16) gives

$$\frac{E_{barrier}(t)}{kT} = \Delta \cdot \left(1 - \frac{H_0}{H_{anis}} \frac{t}{t_0}\right)^2 \tag{A2.3}$$

The substitution of Eq. (A2.3) into (A2.1) and integration gives

$$\ln\left[P_{not}(t)\right] = -\frac{t_0}{\tau_0} \frac{H_{anis}}{H_0} \sqrt{\frac{\pi}{4\Delta}} \left(Erf\left[\sqrt{\Delta}\right] - Erf\left[\sqrt{\Delta}\left(1 - \frac{H_0}{H_{anis}} \frac{t}{t_0}\right)\right]\right) \tag{A2.4}$$

where Erf(t) is the error function $Erf(t) = \frac{2}{\sqrt{\pi}} \int_0^t \exp\left[-x^2\right] \cdot dx$

When the magnetic field is ramped from $H_{min}$ to $H_{max}$, with the rate R, the probability that the magnetization is not switched, is calculated as

$$\ln\left[P_{not}(t)\right] = -\frac{H_{anis}}{\tau_0 \cdot R} \frac{\sqrt{\pi}}{2\sqrt{\Delta}} \left(Erf\left[\sqrt{\Delta}\left(1 - \frac{H_{min}}{H_{anis}}\right)\right] - Erf\left[\sqrt{\Delta}\left(1 - \frac{H_{max}}{H_{anis}}\right)\right]\right) \tag{A2.5}$$

Eqs. (A2.4) and (A2.6) can be used to evaluate $\Delta$ and $\tau_0$ from the measurements of the probability of the magnetization reversal in the ramp magnetic field[32–34]. In order to avoid a systematic error in this method, $H_{anis}$ should be measured from an independent magnetic-static experiment (See Appendix 3).

## Appendix 3

In this Appendix, the different methods for measurement of the parameters of the thermally- activated magnetization switching (TA-switching) are discussed and compared to the proposed method. Only methods, which are based on the Néel model, are discussed. Each measurement method uses some approximations and simplifications, which often lead to a poor measurement precision, a systematic error or even an incorrect result. In the following, common systematic errors of the measurement, statistical analysis and evaluation of parameters of the TA-switching for different measurement methods are discussed. The possible techniques to avoid a systematic error are explained.

The first common error is that an unjustifiably large number of free parameters is used to fit experimental data. It should be noticed that the Néel model is relatively simple and only has two free parameters: the energy barrier $E_{barrier}$ and the rate of interaction $f_{inter}$ of the nanomagnet with the particles (See Appendix 1). Alterternatively, any other pair of free parameters may be used (for example, the coercive field $H_c$ and retention time $\tau_{retention}$ or $M_{eff}$ may be used as one of the two free parameters). However, maximum two parameters should be always used for the data fitting. Additionally, there are parameters, which are related to the magneto-static properties of a nanomagnet. For example, the $\Delta$ is proportional to the anisotropy field $H_{anis}$ (See Eq.(19)) and the volume of the nucleation domain is proportional to the magnetization M of the nanomagnet (See Eq.(17)). Both $H_{anis}$ and M can be measured from an independent magneto-static experiment without the use of any thermally-activated switching measurements. The magnetization M of a ferromagnetic metal can be measured by a magnetometer. $H_{anis}$ can be measured by applying an in-plane magnetic field and monitoring the in-plane component of the magnetization (See main text).

The fact that there are only two free parameters of the Néel model, can be confirmed by the data of Fig.2. A straight line perfectly fits to all experimental data and the line is described by only two free parameters. In the case when condition (9) is not satisfied, the experimental data deviate from the line. From the deviation one more parameter can be evaluated (for example, $\Delta$ or $H_{anis}$). However, a measurement of the deviation from a line of the data of Fig.2 is hard and requires many high-precision measurement points[32,34]. An independent measurement of the $H_{anis}$ from a magnetostatic measurement gives a much better measurement precision of Δ.

The second common error is the assumption that the attempt frequency $f$ is a universal constant of the Néel model and equals to 1 GHz[32,34,35]. It is an unjustified assumption (See Appendix 1). $f$ is proportional to $f_{inter}$ (Eqs.(A1.4),(A1.12)) and it is one of the free parameters of the Néel model. The measurement of $f$ is straightforward (See Eq.(13)). The value of $f$ is material-dependent. There aren't no any facts or experimental evidences proving that the $f$ should be equal to 1 GHz[32,34,35].

The third common error is the incorrect use of statistical measurement and statistical analysis. The following example demonstrates how an incorrect statistical analysis may lead to an incorrect measurement result. The coercive field $H_c$ can be measured as the switching field, at which the magnetization reverses its direction (See Fig.1). Since the $H_c$ depends on the measurement time (See Fig.2), a pulsed magnetic field of an increasing intensity (Fig.8) can be used for the measurement. The field, at which the magnetization is reversed, can be associated with $H_c$ and the measurement time with the pulse duration. The repetition of such measurement gives a distribution of fields at which the magnetization is switched. The average of the distribution gives $H_c$. The parameter Δ can be calculated from the width of the distribution[36]. Such a measurement may have a systematic error due to an incorrect definition of the measurement time. The measured values of $H_c$ and Δ may be substantially different from the correct values. The reason for the error is the following. In the case of Fig. 8, the measurement time does not equal the pulse duration. It is longer and is different for each repeated measurement. This is because, in addition to the one last pulse, at which the magnetization is switched, all preceding pulses contribute to the probability of the magnetization reversal. The measurement time, which includes the contribution of the proceeding pulses, can be calculated from Eq.(12) as

$$t_{switch} = \sum_{i=1}^{n} t_{pulse} \exp\left(-\frac{(H_i - H_c) \cdot M_{eff}}{kT}\right) \quad \text{(A3.1)}$$

where $t_{pulse}$ is the pulse duration and $H_i$ is the magnetic field of each pulse. When Eq.(A3.1) is used in the statistical analysis, the measurements of Fig.2 and Fig.8 give exactly the same values of $H_c$ and $\tau_{retention}$.

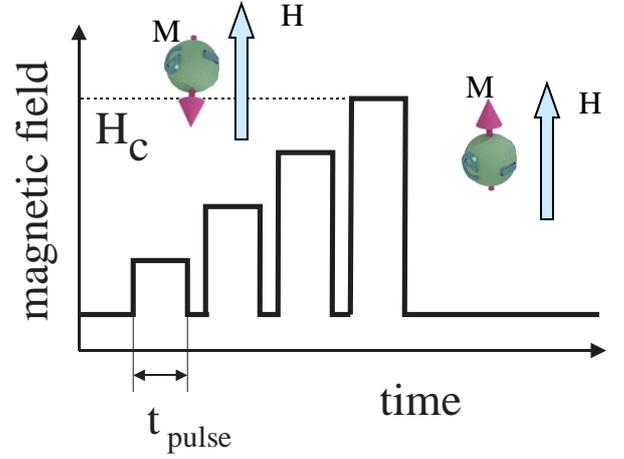

FIG. 8. Method for measurement of $H_c$ using magnetic pulses of gradually- increased amplitude. The magnetic field H (blue arrow) is applied opposite to the magnetization direction (green ball). The field, at which the magnetization is reversed, is assigned to $H_c$.

The forth common error is the use of an insufficient number of the measurements for the statistical analysis. Two free parameters (e.g. $H_c$ and $\tau_{retention}$) for the thermally-activated magnetization reversal can be evaluated using different measurement techniques[34], where a pulsed or linearly-ramp or constant magnetic field is used and the different distributions of the switching time or switching field can be measured. Each method gives the same values of the two free parameters of the thermally-activated switching. However, the number of required repeated measurements to obtain a given measurement precision is different from method to method. For example, in the case of the measurement of Fig.8, only about 30 measurements are required to obtain $H_c$ with a precision better than 1 Oe. However, to obtain the second free parameter (e.g. Δ or $\tau_{retention}$) with even a moderate precision a substantially larger number of measurements is required (~200-500).

There are cases when a measurement of only one parameter of the TA-switching is required. For example, in the case of a VCMA measurement, the gate voltage does not affect the size of the nucleation domain. It reduces the number of the TA-switching parameters, which may be influenced by the gate voltage, to one parameter. For example, the measurement of gate-voltage dependence of only one parameter Hc is sufficient to describe the VCMA effect (See Fig.6). The merit of the measurement of Fig.8 in comparison to the measurement of Fig.2 is a sufficiently shorter total measurement time. As a result, the measurement method of Fig.8 is beneficial for the evaluation of the VCMA effect. In the case of the evaluation of the SOT effect, two parameters of the TA-switching are influenced by the bias current (See Fig.7) and the measurement method of Fig.8 cannot be used.